\documentclass{aa}  
\usepackage{graphicx}
\usepackage{txfonts}
\usepackage{bm}
\usepackage{media9}

\begin{document}

   \title{Stellar streams as gravitational experiments}

   \subtitle{II. Asymmetric tails of globular cluster streams}

   \author{G. F. Thomas\inst{1}, B. Famaey\inst{1}, R. Ibata\inst{1}, F. Renaud\inst{2}, N. F. Martin\inst{1} \and P. Kroupa\inst{3,4}}

   \institute{Universit\'e de Strasbourg, CNRS, Observatoire astronomique de Strasbourg,
		UMR 7550, F-67000 Strasbourg, France\\
              \email{guillaume.thomas@astro.unistra.fr}
              \and
              Department of Physics, University of Surrey, Guildford, GU2 7XH, UK
             \and
              Helmholtz-Institut f\"ur Strahlen-und Kernphysik, Universit\"at Bonn, D-53115 Bonn, Germany
              \and
              Charles University in Prague, Faculty of Mathematics and Physics, Astronomical Institute, CZ-18000 		Praha 8, Czech Republic\\}

   \abstract{Kinematically cold tidal streams of globular clusters (GC) are excellent tracers of the Galactic gravitational potential at moderate Galactocentric distances, and can also be used as probes of the law of gravity on Galactic scales. Here, we compare for the first time the generation of such streams in Newtonian and Milgromian gravity (MOND). We first compute analytical results to investigate the expected shape of the GC gravitational potential in both frameworks, and we then run N-body simulations with the Phantom of Ramses code. We find that the GCs tend to become lopsided in MOND. This is a consequence of the external field effect which breaks the strong equivalence principle. When the GC is filling its tidal radius the lopsidedness generates a strongly asymmetric tidal stream. In Newtonian dynamics, such markedly asymmetric streams can in general only be the consequence of interactions with dark matter subhalos, giant molecular clouds, or interaction with the Galactic bar. In these Newtonian cases, the asymmetry is the consequence of a very large gap in the stream, whilst in MOND it is a true asymmetry. This should thus allow us in the future to distinguish these different scenarios by making deep observations of the environment of the asymmetric stellar stream of Palomar 5. Moreover, our simulations indicate that the high internal velocity dispersion of Palomar 5 for its small stellar mass would be natural in MOND.}


   \maketitle

%-------------------------------------------------------------------
\section{Introduction}
Modified Newtonian Dynamics \citep[MOND,][]{milgrom_1983} postulates that for gravitational accelerations  below $a_0 \approx 10^{-10} {\rm m} \, {\rm s}^{-2}$ the actual gravitational attraction approaches $(g_N a_0)^{1/2}$ where $g_N$ is the usual Newtonian gravitational field generated by baryons. This paradigm provides a natural explanation to the apparent conspiracy between the observed baryonic distribution and the gravitational force in various types of galaxies \citep[see, e. g. ][]{famaey_2012,mcgaugh_2016a,lelli_2016a}. Nevertheless, the successes of the paradigm are mostly limited to galactic scales \citep[for problems at other scales, see, e.g.,][]{skordis_2006,clowe_2006,angus_2008,ibata_2011a,dodelson_2011,clowe_2012,harvey_2015}. This could mean that it is either an effective way to predict how dark matter (DM) is arranged in galaxies, or that it is an effective modification of gravity resulting from a broader theory of the dark sector \citep[e.g.][]{blanchet_2015,berezhiani_2015}. To distinguish between these possibilities, it is absolutely mandatory to check for distinctive predictions of this paradigm in its {\it a priori} domain of validity, i.e. in galaxies. The main reason for the lack of exploration of {\it all} distinctive predictions of MOND on galaxy scales is its intrinsic breaking of the strong equivalence principle, preventing us from treating free-falling systems separately from their host gravitational field. Thus, simulating complex dynamical phenomena such as the disruption of galactic satellites in this framework has been until now hampered by the lack of numerical codes designed to simulate non-symmetrical situations in an N-body context. This has recently changed thanks, notably, to the recent generalizations of the Ramses code \citep{teyssier_2002,lughausen_2015,candlish_2015}. The version we use hereafter and in previous work \citep[e.g.,][]{renaud_2016a,thomas_2017}, was developed by \citet{lughausen_2015}.

Tidal streams of disrupting stellar systems around a host galaxy are very powerful probes of the shape of the gravitational potential of the host, but can also be used as ``leaning tower" experiments to probe gravity on galactic scales. For instance, in the dark matter context, the streams of DM-dominated satellite galaxies could be asymmetric if the weak equivalence principle would be broken for DM particles compared to baryons \citep{kesden_2006a,kesden_2006}. However, this could not be an explanation for asymmetric streams whose progenitors are DM-free globular clusters (GC). The MOND gravitational framework, on the other hand, breaks the {\it strong} equivalence principle, meaning that the internal dynamics of an object is affected by the external field in which it is embedded, even on scales where this external field can be treated as constant. This is true also of general relativistic extensions to MOND like TeVeS \citep[e.g.][]{bekenstein_2004,will_2014} and is known as the {\it External Field Effect} \citep[EFE, see e.g.][]{milgrom_1983,bekenstein_1984,milgrom_2010,mcgaugh_2013,hees_2016a,haghi_2016,mcgaugh_2016,wu_2010,wu_2017}.

Motivated by this, we started a series of papers on the first ever predictions of MOND for tidal streams around the Milky Way (MW). In the first paper of this series \citep{thomas_2017}, we showed that the formation of the Sagittarius stream in MOND is very similar to that in Newtonian dynamics with a spherical DM halo. We also showed that, starting with a dwarf galaxy progenitor sitting on the observed stellar mass-size relation, the remnant was in remarkable agreement with the observations in terms of stellar mass, projected stellar density profile, and stellar velocity dispersion. Nevertheless, our Sagittarius MOND simulation suffered from the same problem as all existing Newtonian models in reproducing the observed stellar velocities in the leading arm. A possible solution to this problem, tentatively proposed in \citet{thomas_2017}, could be a massive squashed hot corona, similar to the triaxial DM halo proposed in Newtonian dynamics, but at the cost of limiting the extent of the stream to distances smaller than 100~kpc for the last 4~Gyr. This problem aside, one important conclusion of this first study, was that, for such a massive progenitor as the Sgr dwarf, the formation of the stream is not affected by the EFE in MOND, so that no breaking of the strong equivalence principle could be probed by such bright and massive streams.

In the present contribution, we set out determine whether the EFE plays a more important role for lower mass progenitors at relatively short Galactocentric distances, namely globular clusters (GC) such as Palomar~5, and whether they could leave a distinctive MOND signature in such kinematically cold streams. In Sect.~2, we summarize the basics of MOND and of the Ramses code patch ({\it Phantom-Of-Ramses}, POR), and review some general predictions for the internal velocity dispersions of GCs. In Sect.~3, we present analytical results on the lopsidedness generated by the EFE for globular cluster models representative of Palomar~5. We present the results of our simulation runs in Sect.~4, and conclude in Sect.~5.

%-------------------------------------------------------------------
\section{MOND}

The POR patch \citep{lughausen_2015} of the Ramses code \citep{teyssier_2002} is based on the quasi-linear formulation of MOND \citep{milgrom_2010}, wherein the generalised Poisson equation reads:
\begin{equation}
\label{eqn:1}
\nabla^2\Phi =  \nabla \cdot \left[\nu\left(\frac{|\nabla \Phi_\mathrm{N} |}{a_0}\right) \nabla \Phi_\mathrm{N} \right] \, ,
\end{equation}
where $a_0$ is Milgrom's acceleration constant and is fixed in the following as $a_0=1.2 \times 10^{-10}$ m.s$^{-2}$, $\Phi$ and $\Phi_\mathrm{N}$ are the MOND and Newtonian potentials respectively. $\nu(x)$ is the interpolation function, which controls the transition from the high acceleration Newtonian regime to the low acceleration deep-MOND regime, such as $\nu(x) =1$ for $x \gg 1$ (Newtonian regime) and $\nu(x)=x^{-1/2}$ for $x \ll 1$ (deep-MOND regime). This Poisson equation is derivable from an action, so that all the standard conservation laws are obeyed \citep{milgrom_2010}. The so-called ``phantom dark matter'' (PDM) density $\rho_\mathrm{ph}$, which can be seen as the virtual (since it does not correspond to real particles) MOND equivalent to the DM contribution in the classical case, is fully defined through Eq.~\ref{eqn:1} once the baryonic distribution $\rho_\mathrm{b}$ (and its associated Newtonian potential $\Phi_\mathrm{N}$) is known, 
\begin{equation}
\rho_\mathrm{ph} (r,z) = \frac{1}{4 \pi G} \nabla \cdot \left[\nabla \Phi_N \ \tilde{\nu}\left(\frac{|\nabla \Phi_N|}{a_0}\right)  \right]\, ,
\label{pdm-form}
\end{equation}
where $\tilde{\nu}(x)=\nu-1$. It is computed at each time-step in POR in order to compute the MOND potential.

A unique prediction of theories like MOND is that the internal dynamics of a satellite system (beyond the usual tidal effects) does not decouple from the external field produced by its host system, an effect which is known as the EFE (see Sect.~1). When the external field dominates over the internal one, it drastically reduces the amount of PDM, and even produces pockets of negative PDM densities at places (see Sect.~3). In general, GCs that have internal accelerations {\it or} external (host galaxy) accelerations larger than $a_0$ should be Newtonian. Interesting cases which should deviate from Newtonian dynamics are those that have both an internal and external acceleration below $a_0$. A list of globular clusters which should be ideal to test MOND has for instance been compiled by \citet{baumgardt_2005}, based on the criterion that the internal acceleration exceeds the external one, so that the EFE can be neglected. This has led to tensions with MOND, especially in the case of Pal~14 \citep{jordi_2009} and Pal~4 \citep{frank_2012}. GCs with a weak external field and extended velocity dispersion profiles, such as NGC~2419, display similar tensions \citep{ibata_2011a}, but see also \citet{gentile_2010} and \citet{sanders_2012}. Note however that, in theories like those of \citet{berezhiani_2015}, such tensions would be expected since MOND would only be an effective description of gravity within a given distance from the host galaxy ($\sim$100~kpc in the case of the MW). However, clusters which are closer to the Galactic center and dominated by the external gravitational field of the MW are also interesting as long as the external field is below $a_0$. This is for instance the case of Pal~5, which is located at a Galactocentric distance of $\sim 18.6$~kpc and experiences an external gravitational field of $\sim 0.5 a_0$, while its internal Newtonian acceleration is well below this. This external field should nevertheless be enough to almost triple the internal velocity dispersion compared to the Newtonian case, as we will see hereafter (Sect.~4). But an even more interesting consequence of the EFE, first noted by \citet{wu_2010}, is that the actual direction of the external and internal gravitational fields being aligned or opposite, depending on the position with respect to the center of the GC leads to the spherical symmetry of the cluster being broken, and its MOND potential should become lopsided. We will delve into this question in more detail in the following section.

\section{Lopsidedness potential of the GCs in MOND}  \label{method}

\begin{figure*}
\centering
  \includegraphics[angle=0, viewport= 0 50 560 520, clip, width=5.8cm]{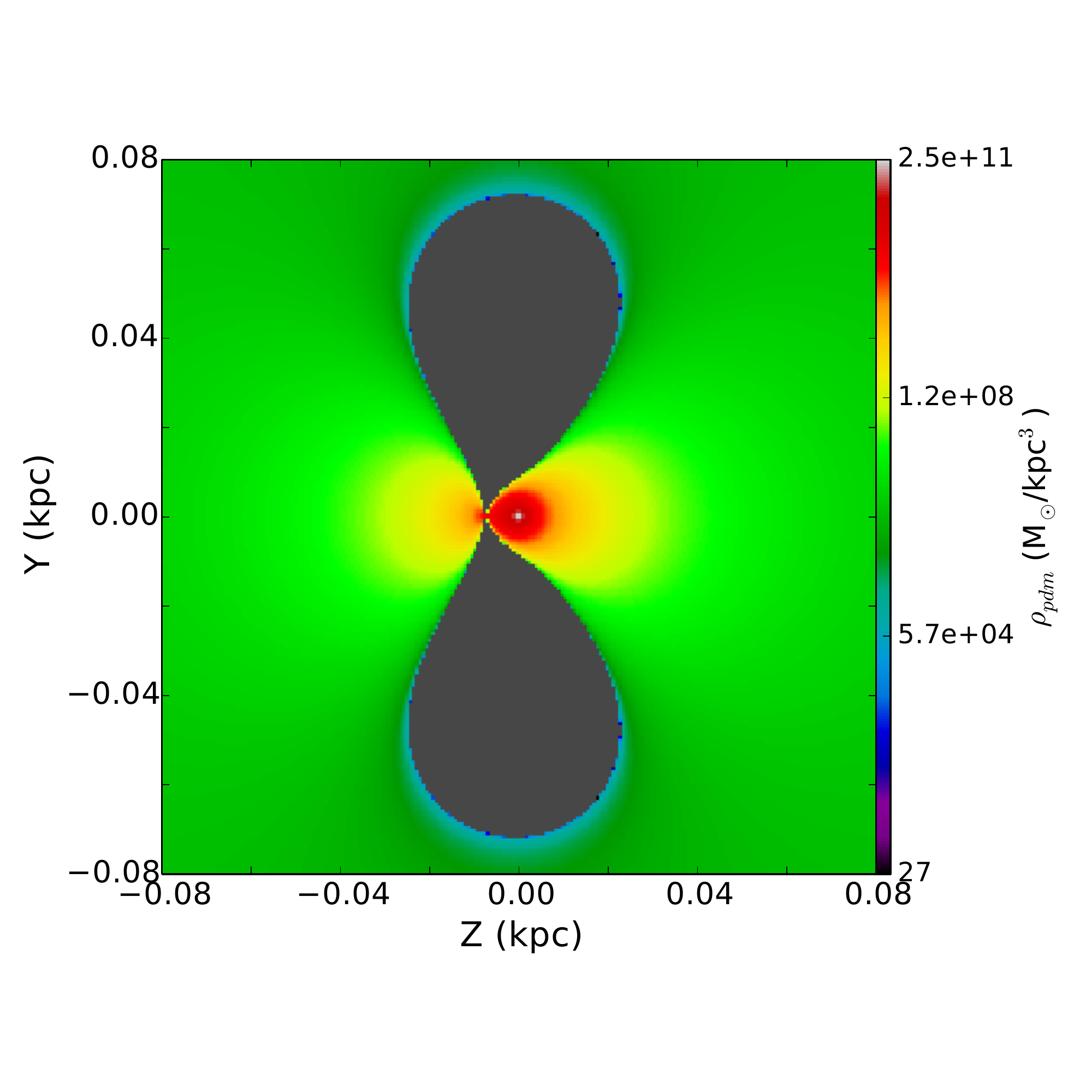}
  \includegraphics[angle=0, viewport= 0 50 560 520, clip, width=5.8cm]{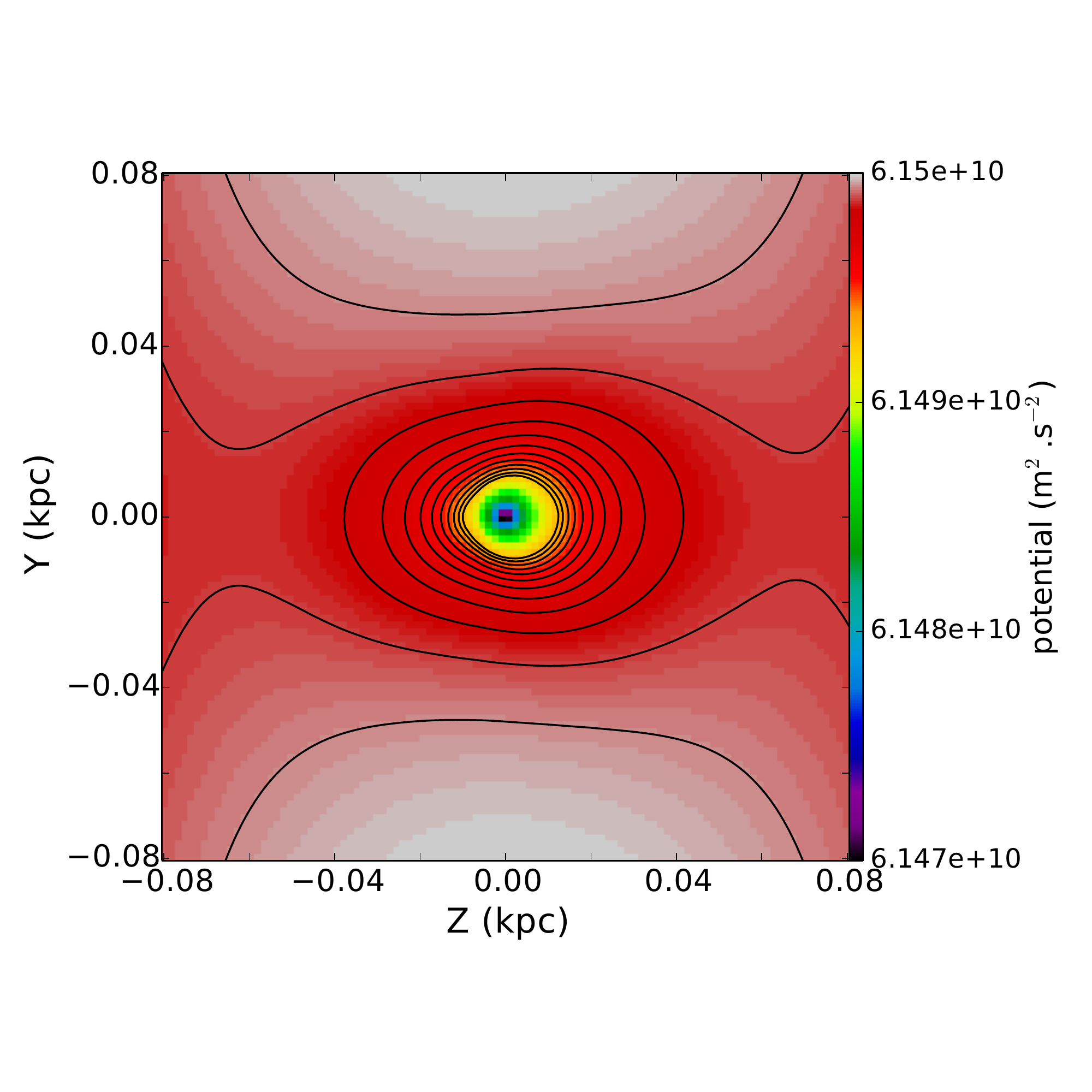}
  \includegraphics[angle=0, viewport= 0 50 560 520, clip, width=5.8cm]{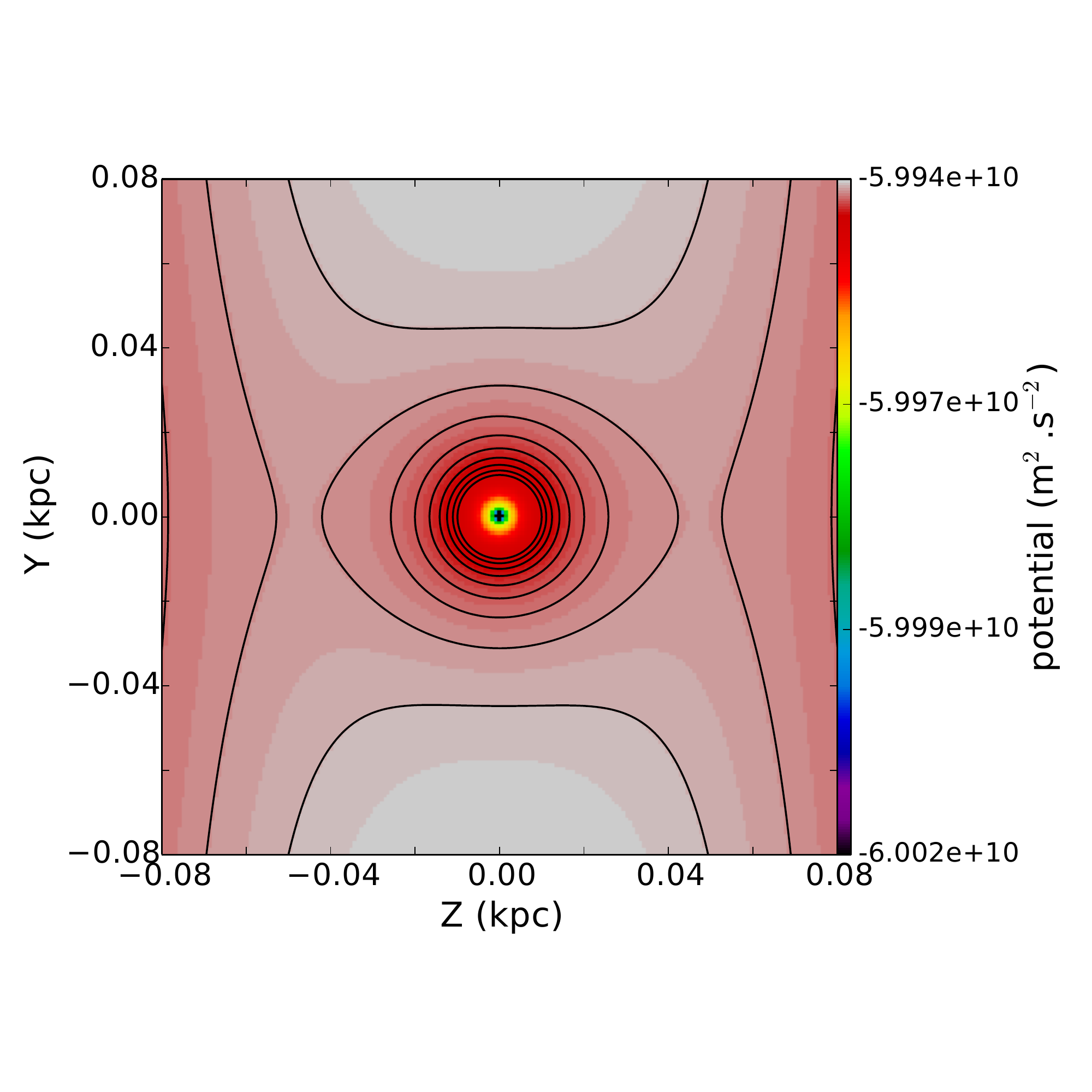}  
  \caption{The left panel displays a cut of the density of PDM around a simple Pal 5-like point-mass ($M_{GC} = 15000$~M$_\odot$) at 14 kpc from the Galactic center ($M_{gal} = 5.56 \times 10^{10}$~M$_\odot$). The Galactic center is at (Z,Y) = (-14,0) kpc on this plot. The grey areas correspond to negative PDM densities. The middle panel displays the corresponding effective potential of the GC. The effective potential of a similar system in Newtonian dynamics is shown in the right panel, where the point mass that represents the MW is significantly more massive than in MOND to account for the dark mass ($M_{gal} = 1.25 \times 10^{11}$~M$_\odot$).}
\label{anal-pdm}
\end{figure*}

As stated in the previous section, the EFE generated by the host galaxy can modify significantly the internal potential of a GC, and thus can have an important impact on the morphology of a tidal stream when the cluster is disrupted.

In this section, we examine a simple case, where two point masses, $M_{GC}$ and $M_{gal}$, represent respectively the GC and the host galaxy. The system is  axisymmetric around the galaxy-GC axis, such that the system can be described in cylindrical coordinates, where the $z$-axis is the axis between the two point masses, pointing outward the host galaxy. The total Newtonian potential, $\Phi_N$, in this configuration can be determined with the following equation: 

\begin{equation}
\Phi_N (r,z) = - \frac{G M_{GC}} {(r^2+z^2)^{1/2}} - \frac{G M_{gal}} { \left[ r^2+(z+D)^2 \right] ^{1/2}}\, ,
\end{equation}
where $D$ is distance between the galaxy and the globular cluster.

The choice of the interpolation function is not crucial here, since the total Newtonian potential in the GCs under consideration is well below $a_0$. But note that the choice of the interpolating function is however important to determine the boost of gravity, and hence of the internal velocity dispersion of the cluster, generated by the external field. Here, following \citet{famaey_2005} and \citet{zhao_2006} for galaxies, we choose an interpolation function of the form : 

\begin{equation}
\tilde{\nu} \left( y \right) = \frac{1}{2} \left( 1+ \frac{4}{y} \right )^{1/2} -\frac{1}{2}
\end{equation}

Since the MOND formalism is non-linear, the PDM density of such a binary system is a composite of the PDM density generated by the host galaxy and the GC when they are isolated, respectively $\rho_{\mathrm{ph},gal}$ and $\rho_{\mathrm{ph},\mathrm{GC}}$, plus a correlating PDM density term $\rho_{ph,cor}$. It is this non-linear term which creates the difference to an equivalent Newtonian dynamics system, where the total density is the sum of the isolated densities. Moreover, depending on the configuration, $\rho_{ph,cor}$ can be negative \citep{milgrom_1986a}, generating a repulsive force. The exact analytical form of the PDM distribution of this binary system is given in Appendix A. It is also plotted in the left panel of Fig.~1, with $M_{GC} = 15000$ M$_\odot$, $M_{gal} = 5.56 \times 10^{10}$ M$_\odot$ and $D = 14$~kpc.

We then used a Newtonian Poisson solver to compute the potential generated by the point mass and the PDM density (from which we subtracted the PDM density generated by the isolated galaxy $\rho_{\mathrm{ph},gal}$), with boundary conditions at $200$~pc around the GC given by \citep{milgrom_1986,famaey_2012} :
\begin{equation}
\Phi_{bound}(x,y,z)= - \frac{G M_{GC} \, \nu_e}{\tilde{r}} \, ,
\end{equation}
with $\nu=\tilde{\nu} + 1$, such that $\nu_e$ is the value of the interpolating function for the value of the host galaxy external field and 
\begin{equation}
\tilde{r} = r_{3D} / \left[ 1+ (L_{Ne}/2) (x^2+y^2)/r_{3D}^2 \right]\, ,
\end{equation}
where $r_{3D}= \sqrt{x^2+y^2+z^2} $ and $L_{Ne} = (\mathrm{d}\, {\rm ln} \nu / \mathrm{d} \,{\rm ln} y )_{y=|\nabla \Phi_N|/a_0}$.

For the plot of phantom dark matter density, we added an $\epsilon$ term to the interpolating function as in Equation~58 of \citet{famaey_2012}, with $\epsilon < 10^{-6}$. The choice of $\epsilon$ below this value had no influence on the resulting potential. The resulting effective potential, assuming a circular orbit around the host galaxy, is plotted on the middle panel of Fig.~\ref{anal-pdm}. It is clear that the effective potential in MOND is lopsided, with a flatening of the isopotentials towards the host galaxy, generated by the lobes of negative PDM. Let us insist here that this case of two point masses is rather extreme, but that a similar lopsidedness is expected in general. Indeed, even if the external field dominates over the internal one everywhere, small negative PDM density lobes will always appear where the projection of the external field along the axis of the internal field is of the same order as the internal field. These lobes will always be asymmetric, as one always expects a stronger MOND boost close to the GC center on the near-side of the host galaxy, because the external and internal fields point in opposite directions, than on the far side. This creates a narrowing of the isopotentials on the near-side, with a stronger MOND boost, making it more difficult for stars at the center of the GC to climb the steeper potential well on the near-side, meaning that we could expect that if a tidal stream is created, it would be asymmetric, with a shorter leading arm (on the Galactic side). For comparison, the effective potential of the same system in Newtonian dynamics is displayed on the right panel of Fig.~\ref{anal-pdm}, with an increased mass for $M_{gal} = 1.25 \times 10^{11}$ due to the contribution of DM around the host galaxy. The choice of parameters for this analytical calculation were inspired by the expected mass and position of the Palomar~5 GC in the MW $\sim$1~Gyr in the past, as we will now see in Sect.~4.

\section{Simulating the Palomar~5 stream in MOND} \label{section-simu}

Motivated by the finding of the previous section, we now investigate whether the expected lopsidedness of the MOND potential of external field dominated globular clusters can have an observable impact on the asymmetry of their tidal tails. For this, we choose the Palomar 5 stream \citep{odenkirchen_2001} as a case study, but the simulations that we are performing hereafter are nevertheless meant to be completely generic: we do not attempt to make a detailed fit of the Palomar 5 tidal stream, since each of our MOND simulations is extremely costly, preventing us from efficiently exploring the parameter space of initial conditions both for the motion of the cluster itself and for its internal original structure. We also point out an important caveat before going further: globular clusters have median two-body relaxation times shorter than their age, meaning that they are collisional systems. Hence, one should ideally model them with collisional direct N-body codes. Such codes are however not available in MOND, and would be extremely challenging to devise. One could imagine simulating many individual two-body and three-body encounters in MOND within a mean gravitational field, and storing the results in a database which would then be used in combination with the POR code, i.e. in a hybrid method, to estimate the effects of the encounters on the dynamics, but this is far beyond the scope of the present contribution. However, in the case of the Pal~5 stream, \citet{dehnen_2004} showed that it can be essentially explained through disk tidal shocking as the cluster passes through the Galactic disk. Therefore, we can model it as a first approximation in a collisionless way. Neglected collisional physics could enhance the mass-loss slightly, but would be isotropic and should not be responsible for any stream asymmetry.

The GC Palomar 5 and its stream have a few interesting and puzzling observational peculiarities. The parameters of the GC are listed in Table~\ref{param_pal5}: in particular, note that the cluster has a very high central velocity dispersion \citep{odenkirchen_2002} for its estimated stellar mass \citep{grillmair_2001,ibata_2017}. This could be linked to a very important contribution of binaries to the velocity dispersion, but as we shall see, a high velocity dispersion is naturally expected in MOND (see also Sect.~2). Beyond this peculiarity of the GC itself, its stream also displays interesting features. In particular, \citet{bernard_2016} used the Pan-STARRS 3$\pi$ Survey to show that the leading arm extends out to $\sim 8 \degr$ from the GC, which is significantly shorter than the trailing arm covering $\sim 16 \degr$ on the sky. Projection effects cannot be the cause of this apparent asymmetry, as we have measured the distance to the stream along most of its length in \citet{ibata_2016}. 
While those CFHT data are significantly deeper and more accurate than Pan-STARRS, they do not probe beyond $6\deg$ in the leading stream. Given the lower depth of Pan-STARRS in those regions, it is not yet clear whether extinction could be the cause of the apparently shorter leading arm \citep[see, e.g.,][]{balbinot_2017}. This will have to be confirmed by further deeper observations with the CFHT. Assuming that a high extinction in the leading arm is not the cause of this apparent asymmetry, it then follows that the stream is also asymmetric in terms of star counts, by a factor $n_{\rm lead}/n_{\rm trail} = 0.72 \pm 0.04$ \citep{ibata_2017}. Apart from external perturbations, Newtonian simulations never produce such asymmetries. They can of course produce small length asymmetries, with a longer or shorter leading arm, due to the different orbital velocities of escaped stars when the orbit of the cluster is eccentric \citep{montuori_2007, mastrobuono-battisti_2012}. However, the Newtonian orbits necessary to reproduce the Pal~5 stream never produce large asymmetries. External perturbations which could cause such asymmetries in Newtonian dynamics include fly-by of compact objects, such as DM sub-haloes or giant molecular clouds \citep{erkal_2016,amorisco_2016} that removed the stars initially present in the leading arm, or a prograde rotating bar in the frame of the stream \citep{pearson_2017}. Of course such effects (apart from those of DM subhaloes) could also exist in MOND, but the key point is that they all create a very large gap in the stream, whilst we would like hereafter to check if the EFE of MOND can on the other hand lead to a true asymmetry. Moreover, this gap can indistinctly affect the leading or trailing tail, with no preference for affecting one over the other. 

\subsection{N-body simulations of the Palomar 5 orbit}

Our MOND simulations are run with the {\it Phantom-Of-Ramses} patch of Ramses \citep{lughausen_2015}. To make faster comparisons, we use the GyrfalcON integrator \citep{dehnen_2000} from the NEMO package \citep{teuben_1995} for the simulations in Newtonian dynamics.
 
As in \citet{thomas_2017}, we model the disk of the MW with the baryonic matter distribution of \citet{dehnen_1998}: a double exponential stellar disk of $3.52 \times 10^{10} M_\odot$ for the thin and thick disk components, with a scale length of 2 kpc and a scale height of 0.3 and 1~kpc respectively. The bulge and the interstellar medium components have respectively a mass of $0.518 \times 10^{10}$ M$_\odot$ and $1.69 \times 10^{10} M_\odot$. In Newtonian dynamics, we added a DM halo following a Navarro-Frenk-White \citep[NFW][]{navarro_1997} profile with an oblateness along the axis perpendicular to the Galactic disk $q_z = 0.94$, a virial mass $M_{200}=1.60 \times 10^{12} M_\odot$, a scale length $r_s=36.5$ kpc, and a concentration $c=5.95$ as determined by \citet{kupper_2015}.  We use the Solar motion of \citet{schonrich_2010} ($U_\odot$, $V_\odot$, $W_\odot$) = ($11.1$, $12.24$, $7.25$) km.s$^{-1}$ in Local Standard of Rest velocities, and a Galactocentric radius of the Sun of $8.5$~kpc. 
 
Contrary to the Sagittarius dwarf spheroidal, which is massive enough and far enough from the Galactic center at apocenter to consider its progenitor as isolated \citep{thomas_2017}, in the case of the Palomar 5 GC, whose apocenter is at $\sim 20$~kpc and which is dominated by the MW external field, we use a Newtonian profile with renormalized gravitational constant $G_{norm} \simeq (a_0/g_{ext}) \times \ G$ \citep{famaey_2012}. Thereby, we modelled the initial progenitor with $10,000$ particles following a Newtonian King profile \citep{king_1966}, where $G$ is replaced  by $G_{norm}$, and we let it orbit the MW for $2$ Gyr. The code is based on Adaptative-Mesh-Refinement (AMR) that increases the resolution of the grid in higher density regions: here we chose a minimum resolution of the AMR grid of 8~kpc and a maximum resolution of 2~pc within the GC and 7 pc within the stream. This is the scale at which our gravity is smoothed. The total size of the grid is 1~Mpc, in order to be able to use spherical point mass-like MONDian boundary conditions.

 \begin{table}
 \centering
  \caption{Current properties of the Pal 5 GC. The bibliographic sources are : $1 =$ \citet{dicriscienzo_2006}, $2 =$ \citet{ibata_2016}, $3 =$ \citet{odenkirchen_2002}, $4 =$ \citet{fritz_2015}, $5 =$ \citet{ibata_2017}.}
  \label{param_pal5}
  \begin{tabular}{@{}ccc@{}}
  \hline
   Parameter & Value & Source  \\
    \hline
   RA & $15^h16^m5.3^s$ & 1 \\
   Dec & $-00\degr 06'41.0"$ & 1 \\
   Distance & $23.5$ kpc & 2 \\
   V$_{rad}$ & $-58.7 \pm 0.2$ km.s$^{-1}$ & 3\\
   $\mu_\alpha$ & -2.3 $\pm$ 0.2 mas.yr$^{-1}$&  4 \\
   $\mu_\delta$ & -2.26 $\pm$ 0.2 mas.yr$^{-1}$&  4 \\
   Mass &  $4297 \pm 98$ M$_\odot$ & 5\\
   r$_t$ &  $0.145 \pm 0.009$ kpc & 5\\
   $\sigma_c$ & $0.9 \pm 0.2$ km.s$^{-1}$& 3\\
\hline
\end{tabular}
\end{table}

We used the current position and velocity of the cluster, listed in Table~\ref{param_pal5}, as the required final phase-space position of the progenitor in the simulation. We allowed the proper-motion free to be within one sigma, and ran only 50 N-body simulations to best match the position of the cluster and stream on the sky. The best choice of proper motion among these 50 simulations was $\mu_\alpha=$-2.24~mas.yr$^{-1}$, $\mu_\delta=$-2.09~mas.yr$^{-1}$. Nevertheless, a simulation with such an orbit still displays a small offset with the observed Pal~5 stream sky-position determined by \citet{ibata_2017} in the standard coordinates ($\xi$, $\eta$), where ($\xi$, $\eta$) = (0, 0) correspond to the center of the GC:
  \begin{equation}
 \left.
  \begin{array}{ l }
\eta_{trailing}(\xi)= 0.211 + 0.768\, \xi -0.0305\, \xi^2 + 0.000845\, \xi^3\\
\\

\eta_{leading}(\xi)= -0.184 + 0.957\, \xi +0.0217\, \xi^2 + 0.00502\, \xi^3 \, .
 \end{array}
  \right.
  \label{position_stream}
\end{equation}
This is the sign that the final proper motions that we used are still slightly wrong, but due to the long time to run one simulation, we assume that this orbit is sufficiently good to be used in this paper for studying the leading-trailing arms asymmetry. We then ran $\sim 50$ additional simulations with different internal structures for the progenitor, in order to find a set of parameters approaching the parameters listed in Table~\ref{param_pal5}, again keeping in mind that a perfect match is not the goal here, as only 50 simulations have been run. Note that we did not try to select specifically asymmetric streams, as the orientation of the stream and the final configuration of the GC remnant were the only measures of merit.

Among these 50 initial progenitor structures, the closest outcome is given by a King model of the progenitor with an initial mass of $M= 2.6 \times 10^4$ M$_\odot$, an initial core radius $r_c= 13$ pc, a ratio between the central potential and the velocity dispersion of $W_0 = 2.25$,  such that $\sigma_c =1.8$~km.s$^{-1}$. The simulation also starts at an initial position such that $G_{norm} \simeq 2.5 G$. This is subsequently going to be called our fiducial MOND model. At the end of this simulation, after 2 Gyr, the remnant of the GC has a tidal radius $r_t = 145$ pc and a mass inside this radius of $M_f = 6500$ M$_\odot$ which is relatively close to the value determined by \citet{ibata_2017} who find a current mass of the progenitor of 4300~M$_\odot$ and a tidal radius $r_t = 145$ pc. Again, a perfect match of the internal structure of the Pal~5 cluster is beyond the scope of the present paper. Moreover, the internal collisional physics which we neglected would only act towards making more stars escape the GC, not less.

\begin{figure}
\centering
  \includegraphics[angle=0, viewport= 71 49 852 708, clip,width=8cm]{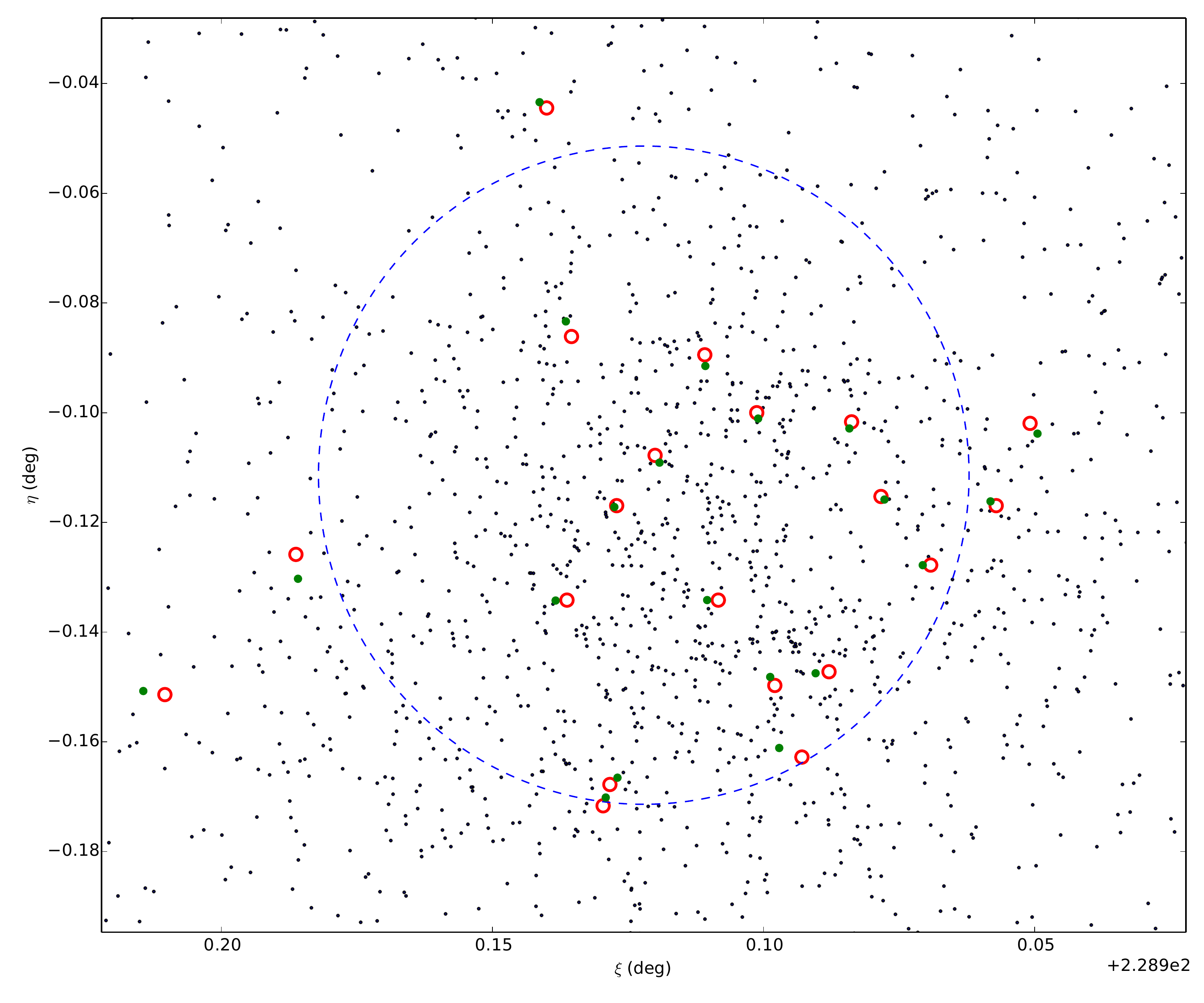}
  \caption{Projection of the central region of GC from our fiducial simulation after 2 Gyr of disruption. The red circles show the position of the stars used by \citet{odenkirchen_2002} to measure the central velocity dispersion and the green dots are the position of the closest particles from our simulation. The blue dashed circle represents the core radius of the Pal 5 GC as measured by \citet{odenkirchen_2002}. The field of view has a size of $12\arcmin \times 10 \arcmin$ ($82$ pc $\times 68$ pc) around the central position of the GC. }
\label{oden}
\end{figure}

To determine the central velocity dispersion of the GC at the end of the simulation, we used two different methods. First, we computed the central velocity dispersion from all the stars inside a core radius of $3.6 \arcmin$ as defined in \citet{odenkirchen_2002}, that leads to a central velocity dispersion of $0.63$ km.s$^{-1}$ in the simulations, which is at 1.35 sigma from the value found by \citet{odenkirchen_2002} of $0.9 \pm 0.2$ km.s$^{-1}$ with the red giant branch stars. For the second method, we used the closest particles to the positions of the 20 stars used by \citet{odenkirchen_2002}, as we show in Fig.~\ref{oden}. With this method, we find a central velocity dispersion of $0.65$ km.s$^{-1}$, 1.25 sigma from the measured value. Such a difference could easily be attributed to the contribution of binary stars to the velocity dispersion. Interestingly, the expected Newtonian velocity dispersion of the remnant GC is much smaller, of the order of 0.2~km.s$^{-1}$, hence more than 3 sigmas away from the measured value. 

\subsection{Results for the stream}

\begin{figure}
\centering
\includemedia[width=5.6cm,windowed=570x575,addresource=./fig_03.mp4,flashvars={source=./fig_03.mp4}]{\includegraphics{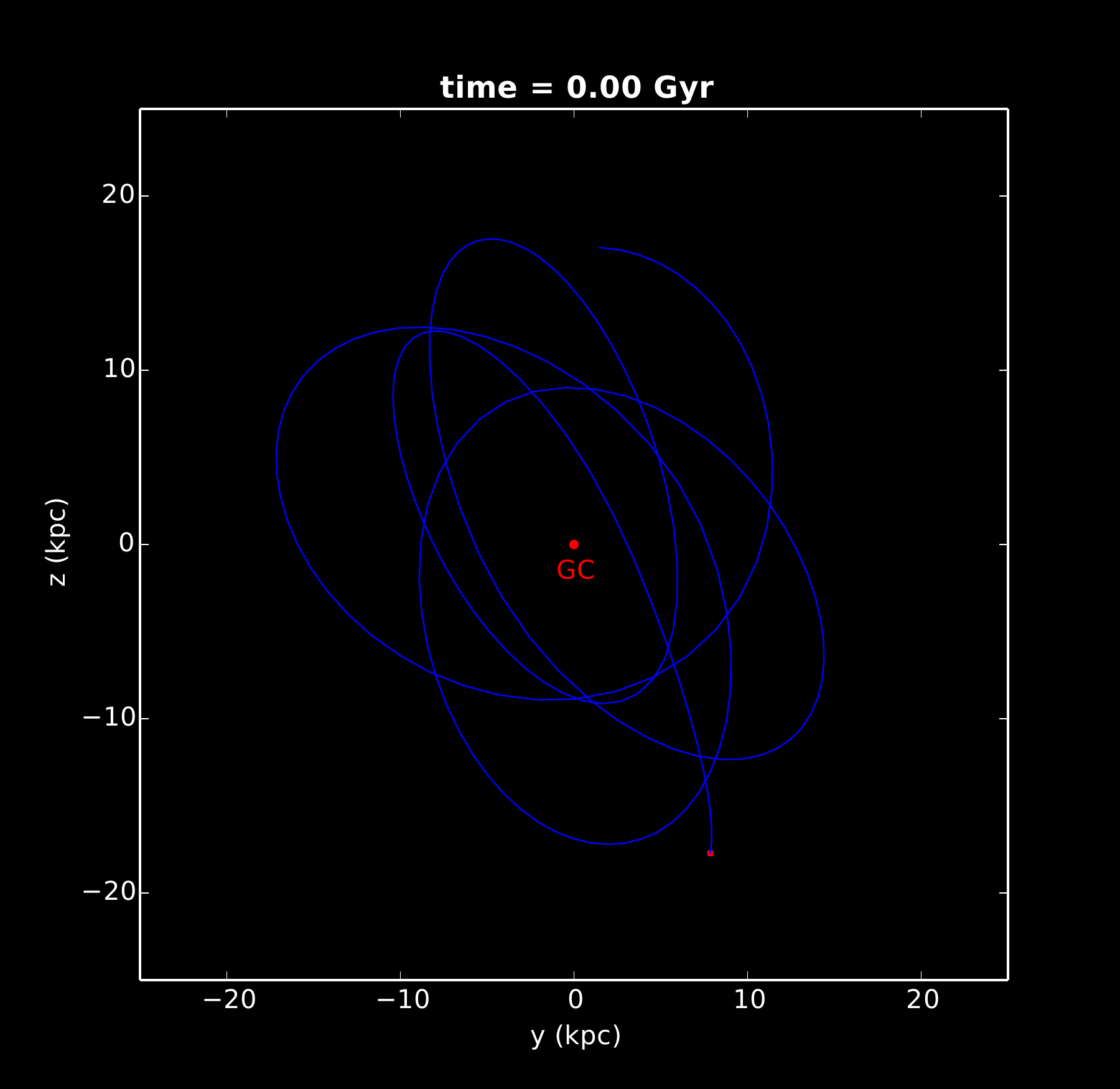}}{VPlayer9.swf}
  \caption{Movie of the fiducial MOND simulation. The particles that are members of the trailing arm at the end of the simulation are color-coded in red. Here the Galactic plane is in the $(x,y)$ plane, and the orbit is shown in the $(y,z)$ plane. The sun is on the x-axis at a distance of $8.5$~kpc.}
 \label{movie}
 \end{figure}

In Fig.~\ref{movie}, we show a movie of our fiducial MOND simulation, where the particles present in the trailing arm at the end of the simulation are color-coded in red. At the end of the simulation, there is a clear asymmetry. Since all the red stars are escaping the progenitor from the outward L2 Lagrange point, it is clear that this leading-trailing arm asymmetry is not the consequence of particles initially present in the leading arm that end up in the trailing arm, but rather the consequence of the shape of the internal potential of the GC prior to disk shocking.

\begin{figure*}
\centering
  \includegraphics[angle=0, viewport= 0 50 560 520, clip, width=7.5cm]{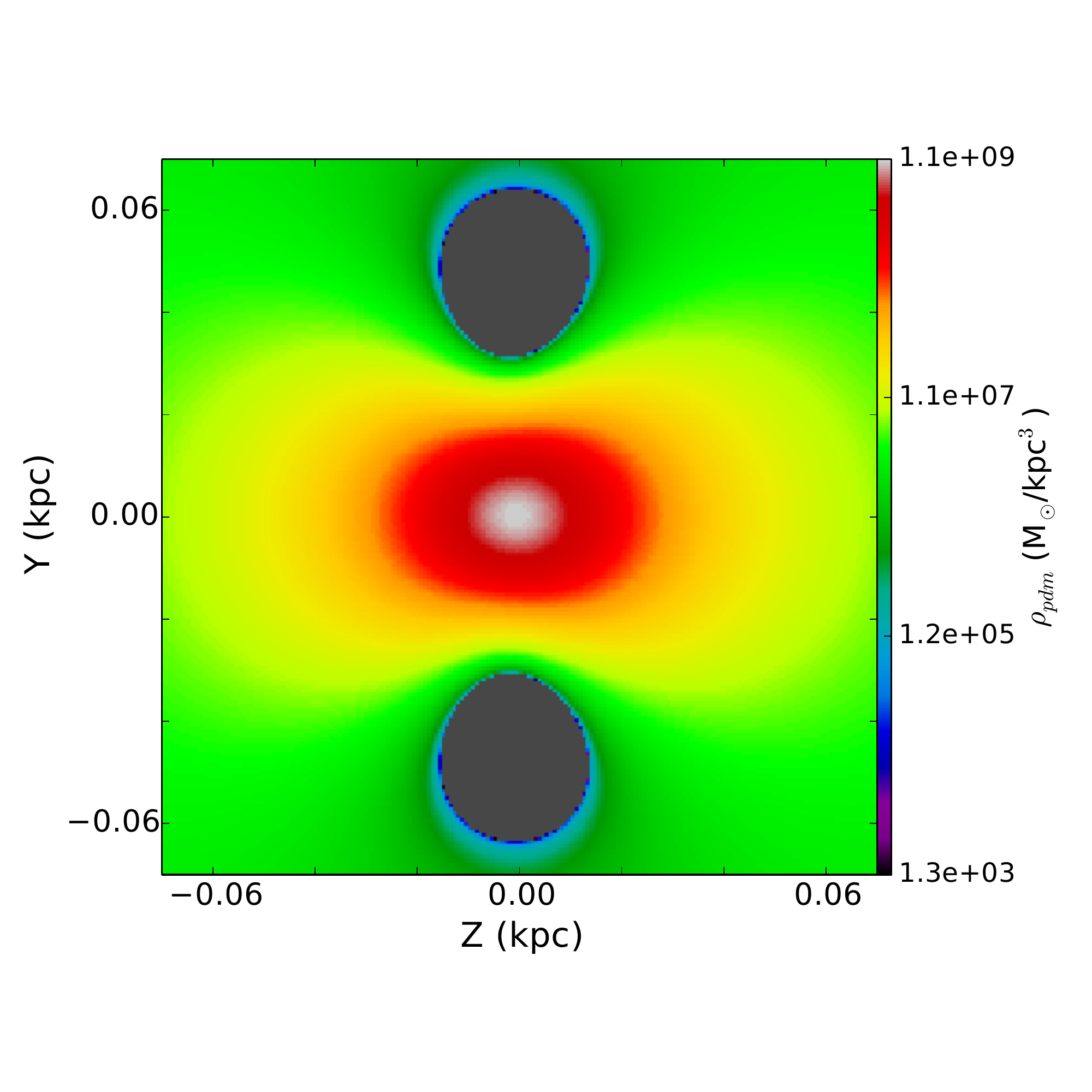}
  \includegraphics[angle=0, viewport= 0 0 578 520, clip, width=6.6cm]{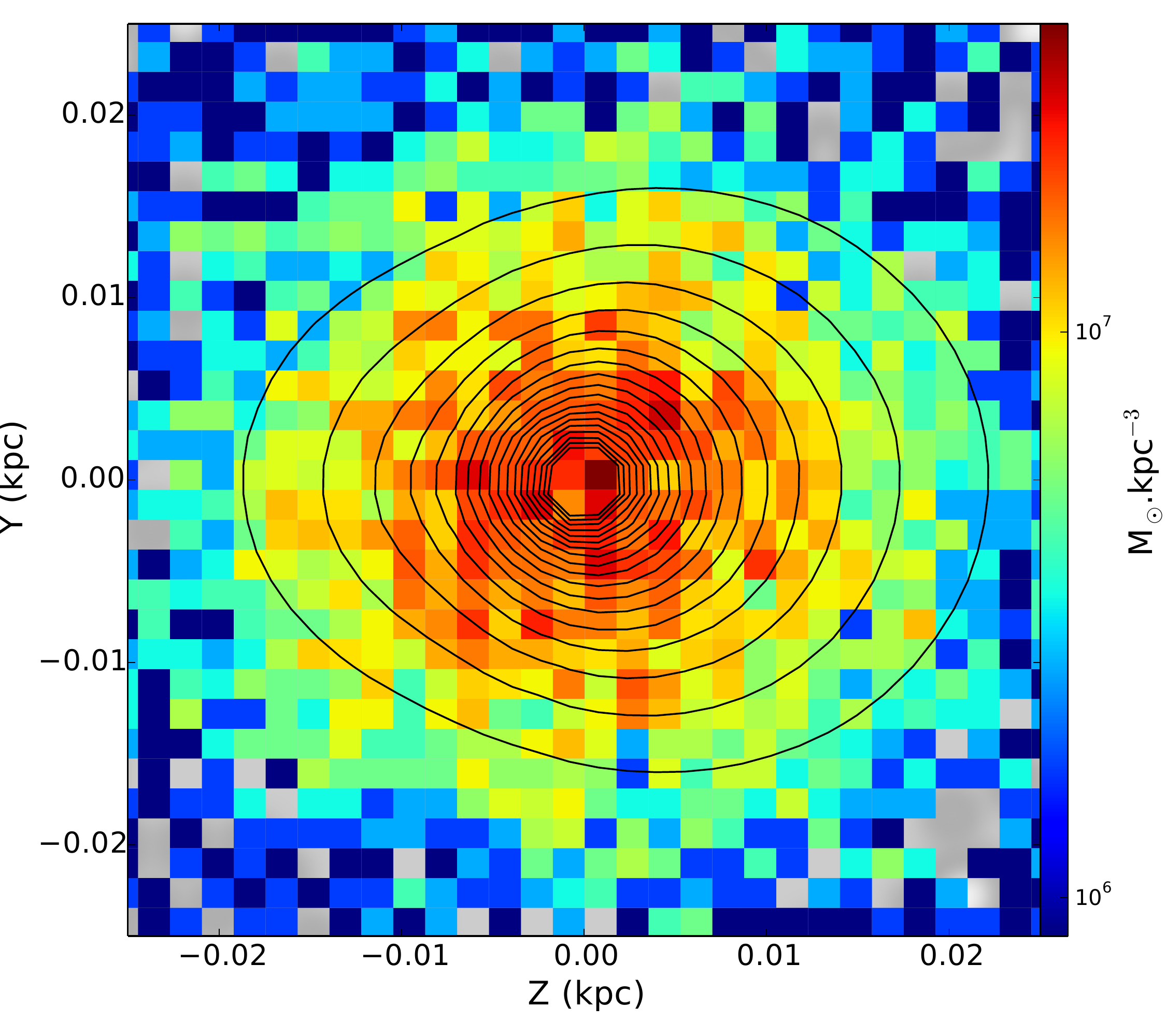} 
  \caption{The left panel displays a cut through the density of PDM around a Pal 5-like Plummer progenitor at 14 kpc from the Galactic center. The Plummer profile has parameters $M=15,000$ M$_\odot$, Plummer radius$= 15$ pc, being similar to the King model of our simulation. The Galactic center is at (Z,Y) = (-14,0) kpc on this plot. The right panel displays the corresponding effective potential of the GC, together with the surface density of the simulated GC particles when the simulated cluster is at the same position as the Plummer model. The lopsidedness is slightly more pronounced in the particles of the simulation than in the isopotentials of the analogous Plummer model.}
\label{partpot}
\end{figure*}

In Fig.~\ref{partpot}, we plot the PDM density, analogous to the simplified point mass model we analyzed in Sect.~3, but for a Plummer model with Plummer radius $=15$~pc and mass$= 15,000$ M$_\odot$, located at a distance of 14~kpc from the Galactic center. This Plummer model is close to our simulated Pal~5 cluster at t=1.1~Gyr in the simulation. We then also plot on Fig.~\ref{partpot} the surface density of our simulated particles projected in the plane containing the Galactic center direction and perpendicular to the tangential velocity, and compare them to the effective potential of the forementioned Plummer model. The lopsidedness of the simulated GC is clear on that plot. This snapshot of the simulation corresponds to the GC being close to reaching its maximum height above the Galactic plane after disk shocking has occured. The lopsidedness of the isopotentials has a double effect: it kinematically heats stars close to the center on the near-side, and implies a smaller number of stars further away from the center on te near-side. Hence when disk shocking occurs, there are fewer stars escaping in the leading arm due to the smaller number of stars in the outskirsts, and stars coming from the inner parts are hotter than on the far-side, thus creating a slightly less massive and more fluffy leading arm compared to the trailing arm. This is also seen in simulations on circular orbits.

\begin{figure*}
\centering
  \includegraphics[angle=0, viewport= 0 0 575 575, clip, width=14.5cm]{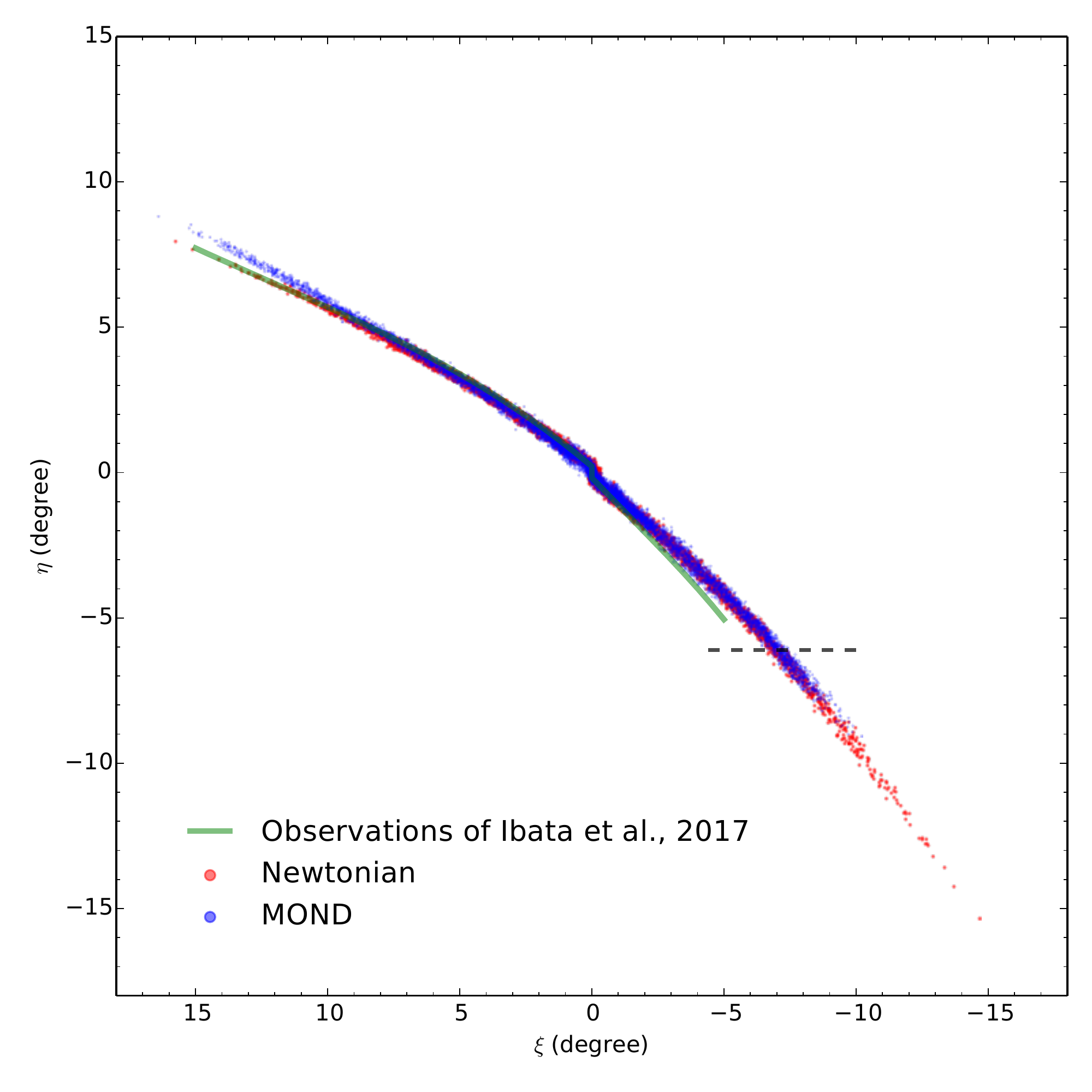}
  \caption{N-body particles of the fiducial MOND simulation (in blue) and of the Newtonian simulation (in red). The green line represents the fit of the center of the stream \citep{ibata_2017} and the dark dashed line shows the limit of detection on the stream in the Pan-STARRS data \citep{bernard_2016}.}
\label{xkieta}
\end{figure*}

\begin{figure*}
\centering
	   \includegraphics[angle=0, width=19cm]{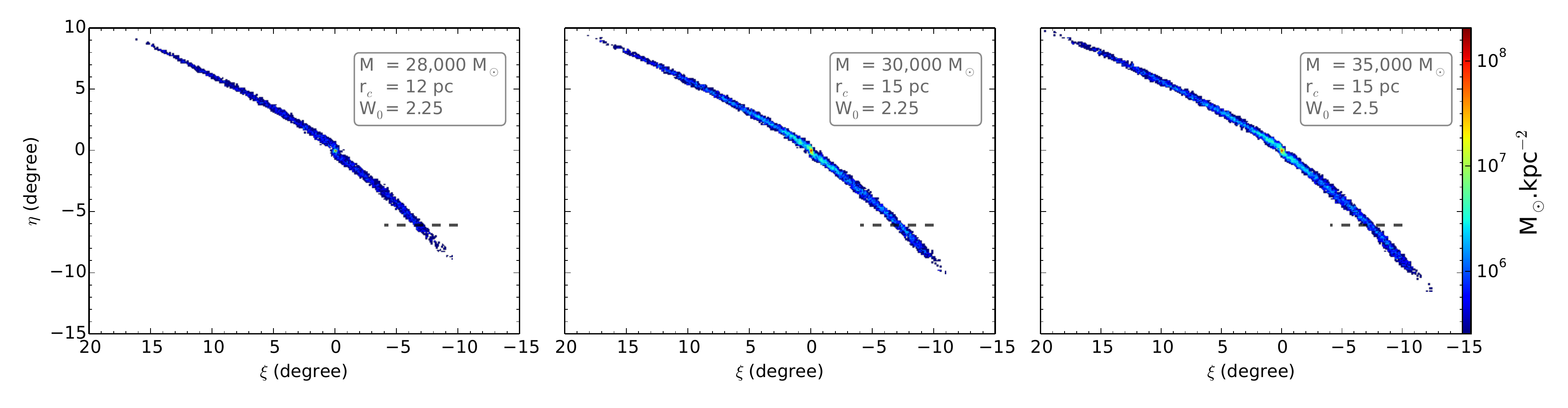}
  \caption{Projection of the streams formed by the disruption of 3 different initial progenitor in MOND. The color  represents the surface density of the stream.}
\label{xkieta_Other}
\end{figure*}

In Fig.~\ref{xkieta}, we show the projection of our fiducial MOND simulation and of a similar simulation in Newtonian dynamics. This Newtonian simulation was not made to reproduce the observed stream and its progenitor but to give a point of comparison between the prediction in the two gravitational frameworks. On this figure, it is clear that in MOND, the leading arm is significantly shorter than the trailing arm at the end of the simulation, which is not the case for the Newtonian simulation. However, the leading arm of the MOND simulation ends at $\eta= 7\degr .5 $, which is $\sim 1\degr.5 $ longer than observed in Pan-STARRS, whose limit is represented by the dashed line on the figure. Nevertheless, the trailing arm in these observations extends only up to a length of $\xi \sim 11\degr$, significantly shorter than measured by \citet{ibata_2016} with deeper observations where the trailing stream can be detected out to $\xi= 15 \degr$. Thus deeper observations of the leading arm should be necessary to determine the length of the leading arm at a similar depth than for the trailing arm. Finally, we note that the radial velocity distribution along the stream at the end of the fiducial MOND simulation is in good agreement with the observed measurements of \citet{odenkirchen_2009} and of \citet{ibata_2017}. 

In the simulation, the leading arm is only slightly less populated than the trailing arm, since the density asymmetry in the outskirts of the GC is very mild. However, since the leading arm is also more fluffy, assuming that the limit of detection of the stream is of 31.5 mag.arcsec$^{-2}$, the {\it detected} leading arm is then actually less populated than the trailing arm by a factor 0.75, close to the value found by \citet{ibata_2017} of $0.72 \pm 0.04$. Indeed, the leading arm has a mean width of $0.2\degr$, whilst the trailing arm has a mean width of only $0.14\degr$.

In our study, we explored 50 different sets of parameters for the initial progenitor, and chose the forementioned fiducial model as a reasonable match to the observations. However, at a more generic level, our 50 simulations allowed us to assess that, in most of the cases, the leading-trailing arm length asymmetry is rather pronounced, with a ratio between the length of the leading and of the trailing arms between 0.5 and 0.8. However, we note that the final asymmetry as well as the dynamical properties of the remnant are very sensitive to the initial conditions: for instance, if the initial core radius is $r_c>20$ pc, the progenitor is completely destroyed after 2 Gyr. As examples, we also show on Fig.~\ref{xkieta_Other} the results of three different initial parameters generating a stream asymmetry. The asymmetry of the first model is consistent with the observed asymmetry however the mass of its remnant is of $16000$ M$_\odot$ and its central velocity dispersion is of $\sigma_c = 1.36$ km.s$^{-1}$, significantly higher than observed. In the  third case, the leading arm is longer than observed by \citet{bernard_2016} and the remnant mass  is of $\sim 10,000$ M$_\odot$ but the velocity dispersion is of $\sigma_c \simeq 0.8$ km.s$^{-1}$ closest to the measurement of \citet{odenkirchen_2002}. 

As in our fiducial model, in these three other MOND simulations, the number of {\it detectable} particles in the leading arm is always lower than in the trailing arm with a ratio between 0.7 and 0.82. This result is very interesting and leads to a clear difference with the generic prediction in Newtonian dynamics, since the number of particles that escape in the leading and in the trailing arm are then similar \citep[e.g.][]{erkal_2016,balbinot_2017}. These differences between the two dynamics is generated by the shape of the effective potential of the cluster since in Newtonian dynamics the effective potential is symmetric to first order, implying that the stars escape in the same way towards the leading and trailing arms \citep[e.g.][]{erkal_2016,balbinot_2017}. 

\section{Conclusion}

In the MOND framework, the effects of the gravitational environment on the internal dynamics of an object are important, beyond the usual tidal effects. Indeed, any MOND-like effective modification of gravity breaks the Strong Equivalence Principle leading to a modification of the internal potential of an object due to the {\it External Field Effect} (EFE). 

In this paper we studied the effect of the EFE on a Palomar~5-like GC embedded in a MW-like gravitational field. We showed  how asymmetric lobes of negative ``PDM" density are then generated in MOND, leading to a lopsided effective potential for the GC.

We have then run N-body simulations in MOND, inspired by the Palomar~5 case, but not intending to produce a close fit, in order to study generic differences between the tidal stream generation in MOND and Newtonian dynamics. We found that the GC is disrupted due to tidal disk shocking, but compared to its Newtonian counterpart, the leading arm of the stream is significantly shorter, slightly less massive, and significantly more fluffy than the trailing arm, due to the asymmetry of the effective potential of the cluster. Assuming that the limit of detection of the stream is of 31.5 mag.arcsec$^{-2}$, the {\it detected} leading arm is then actually less populated than the trailing arm by a factor between 0.7 and 0.82, depending on the exact MOND model of the progenitor GC. Moreover, we note that the velocity dispersion of the remnant GC is always much higher than in the Newtonian counterpart, thereby bringing the simulated velocity dispersion closer to observations than in Newtonian dynamics for the low stellar mass of the cluster. 

In the Newtonian framework, such a large leading - trailing arm asymmetry, which is observed for the Palomar 5 stream in the Pan-STARRS data (provided it is not due to extinction),  can only be the consequence of interactions with dark matter subhalos \citep{erkal_2016}, giant molecular clouds \citep{amorisco_2016}, or interaction with a prograde Galactic bar. In all these Newtonian cases, the asymmetry is the consequence of a very large gap, due to the disruption of the pre-existing stream. On the other hand, in MOND it is a true asymmetry in the outflow making up the stream. This should thus allow us in the future to distinguish these different scenarios by making deep observations of the environment of the asymmetric stellar stream of Palomar 5. Moreover, in MOND, such asymmetries should only affect the leading arms of streams, while other causes would not lead to such a systematic behaviour, so finding other asymmetric streams would also be an important test.

In the future, we will also investigate whether the lopsided potential can modify significantly the shape of GCs in a relevant range of distances from their host galaxy, close enough for the external field to be significant, and far enough not to be destroyed by tidal effects. It will also be interesting to ascertain whether a similar effect occurs in other streams formed by the disruption of a GC. To rule out MOND we would want to study streams that show no such asymmetry, which could be the case for, e.g., the NGC 5466 stream \citep{grillmair_2006c}. However, an important conclusion of our present investigations is that the final asymmetry as well as the dynamical properties of the remnant are very sensitive to the orbit and initial conditions for the progenitor, so that no conclusion can be reached without modelling each different stream separately. Other potentially interesting streams for further studies are those recently discovered around Eridanus, Palomar 15 \citep{myeong_2017} or the PS-1 C stream around the Balbinot 1 GC \citep{bernard_2016,balbinot_2013}.

\section*{Acknowledgements}
The authors acknowlede insightful comments from Moti Milgrom, Paola Di Matteo, Francoise Combes, Edouard Bernard, Gary Mamon, and an anonymous referee. FR acknowledges support from the European Research Council through grant ERC-StG-335936.

  \bibliographystyle{aa}
  \bibliography{./biblio}

\onecolumn

\appendix
\section{Analytic calculation of the phantom dark matter density for two point masses}

In this Appendix, we detail the calculation of the phantom dark matter density generated by the binary system composed of two point masses $M_{GC}$ for the globular cluster and $M_{gal}$ for the host galaxy that is located a distance D away. The coordinates are cylindrical $(r,z)$ centred on the center of the globular cluster where the $z$-axis is pointing outward the host galaxy.

As written in Section \ref{method} the PDM density is linked to the total Newtonian potential by the non linear equation (\ref{pdm-form}). In this system, the total Newtonian potential is given by :

\begin{equation}
\Phi_N (r,z) = - \frac{G M_{GC}} {(r^2+z^2)^{1/2}} - \frac{G M_{gal}} { \left[ r^2+(z+D)^2 \right] ^{1/2}}
\end{equation}

Thus the derivative of this potential, that corresponds to the inverse of the Newtonian acceleration, can be decomposed in two components, one along the $r$-axis and one along the $z$-axis: 

\begin{equation}
\begin{split}
&\nabla \Phi_{N r} =  G \,r \biggl[  M_{GC} \left( r^2 + z^2  \right)^{-3/2} + M_{gal} \left( r^2 + (z+D)^2  \right)^{-3/2}\biggr]  \, ,\\
&\nabla \Phi_{N z} =  G \,  M_{GC} \,z \, \left( r^2 + z^2  \right)^{-3/2} + G  M_{gal} \, (z+D) \left( r^2 + (z+D)^2  \right)^{-3/2} \, .
\end{split}
\end{equation}

The norm of the last derivative is equal to :
\begin{equation}
\begin{split}
|\nabla \Phi_N| \,(r,z) =  G & \bigl[  M_{GC}^2 \left( r^2 + z^2  \right)^{-3} \left( r^2 + z^2\right) \\
& +  M_{gal}^2 \left( r^2 + (z+D)^2  \right)^{-2}  \\
& + 2\,  M_{GC}\, M_{gal} \left( r^2 + z^2  \right)^{-3/2} \left( r^2 + (z+D)^2 \right)^{-3/2}  \left( r^2+z\, (z+D) \right) \bigr]^{1/2} \, .\\
\end{split}
\end{equation}

The partial derivative of these quantities, needed to calculate $\rho_\mathrm{ph}$ can also be split in two components :

\begin{equation}
\begin{split}
\frac{\partial |\nabla \Phi_N|}{\partial r} = \frac{1}{2} |\nabla \Phi_N| ^{-1} G\times & \biggl\lbrace  2 \, M_{GC}^2\, r \, \left(r^2+z^2\right)^{-3} - 6\, M_{GC}^2\, r\, \left(r^2+z^2\right)^{-4}\left (r^2+z^2\right) - 4\,M_{gal}^2\,r \, \left(r^2+(z+D)^2\right)^{-3}\, ,\\
&  + 4\,M_{GC} \, M_{gal} \,  r  \, \left(r^2+z^2\right)^{-3/2} \left(r^2+(z+D)^2\right)^{-3/2} \\ 
& - 6\,M_{GC} \, M_{gal} \,r \, \left(r^2+z\,(z+D)\right)\\
&  \times \biggl[ \left(r^2+z^2\right)^{-5/2} \left(r^2+(z+D)^2\right)^{-3/2} + \left(r^2+z^2\right)^{-3/2} \left(r^2+(z+D)^2\right)^{-5/2}\biggr] \biggr\rbrace \, .\\
\end{split}
\end{equation}

\begin{equation}
\begin{split}
\frac{\partial |\nabla \Phi_N|}{\partial z} = \frac{1}{2} |\nabla \Phi_N| ^{-1} G\times & \biggl\lbrace  2 \, M_{GC}^2\, z \, \left(r^2+z^2\right)^{-3} - 6\, M_{GC}^2\, z\, \left(r^2+z^2\right)^{-4} \left(r^2+z^2) - 4\,M_{gal}^2\,(z+D) \, \left(r^2+(z+D)^2\right)^{-3} \right. \\
&  + 2\,M_{GC} \, M_{gal} \,  \left(2\,z +D\right)  \, \left(r^2+z^2\right)^{-3/2} \left(r^2+(z+D)^2\right)^{-3/2} \\ 
& - 6\,M_{GC} \, M_{gal} \,z \,\left(r^2+z\,(z+D)\right) \, \left(r^2+z^2\right)^{-5/2} \left(r^2+(z+D)^2\right)^{-3/2}\\
& \biggl. - 6\,M_{GC} \, M_{gal} \,(z+D) \, \left(r^2+z\,(z+D)\right) \, \left(r^2+z^2\right)^{-3/2} \left(r^2+(z+D)^2\right)^{-5/2} \biggr\rbrace
\end{split}
\end{equation}

The second derivatives of the potential are expressed along $z$ and $r$, the cross term not being necessary to calculate  $\rho_\mathrm{ph}$ :

\begin{equation}
\begin{split}
\frac{\partial\, \nabla \Phi_{N r }}{\partial r} = & \,  G \left[ M_{GC} \left( r^2 + z^2  \right)^{-3/2} + M_{gal}  \left( r^2 + (z+D)^2 \right)^{-3/2} \right]\\
& - 3 \, G \, r^2 \left[ M_{GC} \left( r^2 + z^2  \right)^{-5/2} + M_{gal}  \left( r^2 + (z+D)^2 \right)^{-5/2} \right] \, ,\\
\end{split}
\end{equation}

\begin{equation}
\begin{split}
\frac{\partial\, \nabla \Phi_{N z }}{\partial z} = & \,  G \biggl[ M_{GC} \left( r^2 + z^2  \right)^{-3/2} + M_{gal}  \left( r^2 + (z+D)^2 \right)^{-3/2} \biggr]\\
& - 3 \, G  \biggl[ M_{GC} \, z^2 \left( r^2 + z^2  \right)^{-5/2} + M_{gal} (z+D)^2 \left( r^2 + (z+D)^2 \right)^{-5/2} \biggr]\, .\\
\end{split}
\end{equation}

Finally, the density of PDM $\rho_\mathrm{ph}$ can be calculated analytically from the previous quantities:

\begin{equation}
\begin{split}
\rho_{\mathrm{ph}} (r,z) = \frac{1}{4 \pi G} \biggl\lbrace & \frac{1}{2 \, r} \left( 1 + \frac{4 \, a_0}{ |\nabla \Phi_N|} \right)^{1/2} \nabla \Phi_{N r} + \frac{1}{2} \left( 1 + \frac{4 \, a_0}{ |\nabla \Phi_N|} \right)^{1/2} \frac{\partial\, \nabla \Phi_{N r }}{\partial r} \\
& \\
& - a_0\left( 1 + \frac{4 \, a_0}{ |\nabla \Phi_N|} \right)^{-1/2} |\nabla \Phi_N|^{-2}  \; \nabla \Phi_{N r} \, \frac{\partial |\nabla \Phi_N|}{\partial r}\\
&\\
& - \frac{\nabla \Phi_{N r}}{2 \, r } - \frac{1}{2} \frac{\partial\, \nabla \Phi_{N r }}{\partial r} - \frac{1}{2} \frac{\partial\, \nabla \Phi_{N z }}{\partial z} 
& \\
&+ \frac{1}{2} \left( 1 + \frac{4 \, a_0}{ |\nabla \Phi_N|} \right)^{1/2} \frac{\partial\, \nabla \Phi_{N z }}{\partial z} \\
& \\
&  - a_0\left( 1 + \frac{4 \, a_0}{ |\nabla \Phi_N|} \right)^{-1/2} |\nabla \Phi_N|^{-2}  \; \nabla \Phi_{N z} \, \frac{\partial |\nabla \Phi_N|}{\partial z} \biggr\rbrace\, .\\
\end{split}
\end{equation}

\end{document}